\documentclass[useAMS]{mn2e}
\usepackage{epsfig}

\title[Unified dark matter: constraints from galaxies and clusters ]{Unified dark matter: constraints from galaxies and clusters}
\author[Amr El-Zant]{Amr A. El-Zant\\
Centre for Theoretical Physics, The British University in Egypt, Sherouk City 11837, Cairo, Egypt\\}

\begin{document}

\date{Received...; in original form....}

\pagerange{\pageref{firstpage}--\pageref{lastpage}} \pubyear{2002}

\maketitle

\label{firstpage}

\begin{abstract}
Unified dark matter models are appealing in that they describe the dark sector in terms of a single  component. 
They however face problems when attempting to account for structure formation: in the linear regime,   
density fluctuations  can become Jeans stable and oscillate rather than collapse, though it is 
possible that this difficulty may be circumvented  by invoking nonlinear clustering. Here we examine the  behaviour in the fully nonlinear regime,  of collapsed objects that should mimic
standard dark matter haloes. It is shown that the pressure gradient associated with the unified dark matter fluid
should be significant in the outer parts of galaxies and clusters, and its effects obervable. In this case, 
no flat or falling rotation curve is possible for any (barotropic) equation of state with associated sound 
speed decreasing with density (a necessary condition if the fluid is to behave as pressureless matter   
at high density). The associated density profile is therefore also incompatible with that inferred 
in the outer part of clusters.  For the prototypical case of the generalised Chaplygin gas, it is 
shown that this limits the values of the equation of state index $\alpha$ that are compatible with observations
to $\alpha \la 0.0001$ or $\alpha \ga 2$.  This is in line from what is deduced from linear analysis. 
More generally, from the expected properties of dark matter haloes,  constraints on the sound speed 
are derived. For the particular case of the generalised Chaplygin gas, 
this further constrains the index to $\alpha \la 10^{-9}$ or $\alpha \ga 6.7$. For a unified dark matter fluid
to mimic dark halo properties, therefore, it needs to have an equation of state such that the pressure gradients 
are either minimal  or which decrease fast enough so as to be negligible at densities characteristic of the 
outer parts of haloes.

\end{abstract}


\section{Introduction}

Barring significant modifications to known physical law, compelling evidence suggests 
that most  of our universe  is composed of missing dark energy and matter.
The dark energy component is compatible with a cosmological constant $\Lambda$, 
which  effectively contributes a negative  pressure,  equal to the energy density in magnitude 
but reversed sign ($p = - \rho$).  Dark matter on the other 
hand generally needs to be  'cold' (or at least warm), in the sense of having small enough 
streaming length at matter and radiation
equality, so as to satisfy  structure formation constraints, including Lyman-alpha bounds
(e.g., Viel et. al. 2013).  The cold dark matter (CDM) 
scenario has developed over three decades into a remarkably successful model for structure formation, 
despite some problems on galactic scales (e.g., Frenk \& White 2012 for a review). 

 Despite the successes of the $\Lambda$CDM cosmology, the origin of the missing components remains 
obscure. In the case of dark energy,  particularly perplexing is the unexpectedly small value
and its relatively recent domination of the energy budget of our universe. 
These issues have triggered much investigation, seeking a myriad of 
alternatives to the cosmological 
constant as the driver of late cosmic acceleration 
(e.g., Li et. al. 2011; Yoo \& Watanabe 2012).     

Dark energy and dark matter need not form separate components.
 In particular, if  a single unified dark fluid obeys an equation 
of state such the magnitude of its (negative) pressure decreases
rapidly enough with density, then it may behave as CDM at high densities and much 
like $\Lambda$ at low densities.
This contention is appealing and plausible in principle, given the observation that the negative pressure
contribution to the content of the universe only becomes important in recent cosmological epochs, 
when the total energy density  has been significantly diluted by expansion; and  
that  the bulk of structure formation, including collapse leading to gravitationally bound structures,  
took place before dark energy became dominant.  CDM-like hierarchical structure formation 
may then proceed naturally, as in $\Lambda$CDM 
prior to $\Lambda$ domination. In this picture, the bound structures, thus formed, remain at high density, impervious to 
later cosmological density dilution, when the unified dark matter gas acts as dark energy on large scales, 
playing the role of $\Lambda$. 

Most extensively studied among unified dark matter models has been the generalised Chaplygin gas
cosmology (Kamenshchik,  Moschella \& Pasquier 2001;  Bento, Bertomlami \& Sen 2002). 
Here, the relevant fluid is endowed with the  equation of state~\footnote{Physical units are used here instead of the 
customary $c=1$, as a principal concern will be to compare typical speeds in galaxies with sound speeds in a 
unified  dark matter fluid.}
\begin{equation}
p = - \frac{A c^2}{\rho^\alpha}.
\label{eq:state}
\end{equation}
The positive parameters $A$ and  $\alpha$ can be adjusted so that the associated 
solutions of the Friedman equations  match the expansion history of the universe, as 
revealed by  supernovae type Ia data. Park et. al. (2010) find a preferred value of $\alpha = 0.997$. 
This is remarkable, as this original Chaplygin gas cosmological model (with $\alpha = 1$) 
is particularly well motivated from a  theoretical point of view 
(Gorini et al. 2006; Jackiw 2000). However, this value of $\alpha$ 
turns out to be in tension with the estimated age 
of the universe, and  in catastrophic contradiction with probes of structure formation.
Indeed, it has long been known that the non-zero pressure  associated with unified dark matter 
may significantly stem the growth of structure: perturbations can become 
Jeans stable; rather than collapse they oscillate, 
imprinting unobserved  signatures on the matter density fluctuation power spectrum 
(Sandvik et al. 2004).  Including the baryonic component in the analysis  of the growth of fluctuations ameliorates the situation somewhat (Be{\c c}a et al. 2003). 
Nevertheless, values of  $\alpha$ very close to zero 
or greater than one (when sound speeds can become superluminal) 
are still favoured (Gorini et. al. 2008a: Fabris et. al. 2008; Fabris, Velten \& Zimdahl 2010).
A final apparent {\em coup de grace} comes from
combining  supernovae data with  matter power spectrum constraints, as well as cosmic
microwave background and baryon acoustic oscillations data. This seems to conclusively 
constrain $\alpha$ to very small values near zero, with the implication that the ensuing model 
is virtually indistinguishable from $\Lambda$CDM (e.g. Park et. al.  2010; Wang et. al. 2013).

The above constraints were deduced by considering linear growth of structure from adiabatic 
perturbations. Non-adiabatic perturbations (e.g., Reis et. al. 2003; Zimdahl \& Fabris 2005; Billic, Tupper \& Viollier 2007) and nonlinearities were therefore considered. 
It is not clear whether initially non-adiabatic perturbations would remain impervious to pressure forces 
into the nonlinear regime, and thus whether invoking these solves
the problem of structure formation in unified dark matter models. 
Nonlinear collapse (including pressure forces) was investigated by Billic et. al. (2004), 
starting from a homogeneous background and using a simple collapse model 
coupled to  the Press-Schecter formalism. They found that 
only a small fraction of initial density perturbations lead to collapsed structures akin to 
dark matter haloes.  Their model  is nevertheless open to refinement.

 Another  crucial   issue  concerns the fact that collapsed unified dark matter structures (that form bound CDM-like haloes), 
'decouple' from the pressure budget, they therefore
do not contribute to the dark energy content (Billic et. al. 2003; Avelino et. al. 2004; Avelino, Bolejko \& Lewis 2014). 
This component, already collapsed into bound systems,  behaves as pressureless  matter, and can thus cluster accordingly. 
Clustering  also affects the background evolution (which is driven by a modified effective pressure),
and it has been suggested that the resulting effective equation of state could allow 
Chaplygin gas cosmologies to be compatible with available observations
for all relevant values of the index $\alpha$, provided the clustering is efficient enough. 
This is a powerful argument, though accurately evaluating the effective equation of state
would require  detailed understanding  of Chaplygin gas hydrodynamics as the fluid
forms structures in a cosmological context, or at least much more detailed modelling 
of the collapse and clustering process than has been hitherto available.    

Nevertheless,  whatever the intermediate evolution, if the final quasi-equilibrium 'haloes' are to be indistinguishable 
from their largely successful CDM counterparts, it is required that they have similar density profiles in the outer
regions. For this to happen, the pressure gradients associated with the Chaplygin gas should either be negligible or 
somehow collude as to 
result in an empirically viable density profile, at least up to radii where direct  discriminants 
(such as rotation curves of galaxies and X-ray emissions from clusters) are available.    
These requirements may not be easily satisfied, as the average density
of CDM haloes within their virial radii is only $\sim 200$ times more dense than the average density of the universe
(by definition) and is smaller in the outer regions; pressure gradients could therefore 
be palpable in the outer parts of galaxies.   
On the other hand, the possibility that 
significant pressure gradients could  collude so as to lead to profiles similar to those of CDM haloes is
unlikely, given that -- as opposed to the case of collisionless CDM -- the existence of definite equation of state
ensures a unique solution to the hydrostatic equilibrium (and associated density profile), given the boundary conditions.

In this context, we consider the effect of pressure gradients on the outskirts of spherical 
unified dark matter haloes in hydrostatic equilibrium, showing that the effect of such 
gradients should be observable if present. In the case of the generalised Chaplygin gas, this  
restricts plausible values of $\alpha$.  
More generally, we provide constraints on possible values of the sound speed 
(and therefore equation of state).
Section~\ref{sec:equi}  
gives a brief introduction to the basic properties of Chaplygin gas, then discusses its self-gravitating 
hydrostatic equilibrium and associated density distribution; more general properties 
of any fluid with sound speed increasing with decreasing density are discussed in Appendix~B. 
The transition from CDM-like behaviour 
is  tackled in Section~\ref{sec:trans}, the resulting outer rotation
 curves are then 
compared with their observed counterparts. In Section~\ref{sec:clus} further, and more 
quantitative, constraints on plausible values of the generalised Chaplygin index $\alpha$ are obtained by comparing 
the density profiles of Chaplygin haloes with observationally inferred cluster halo profiles. In Section~\ref{sec:gen}
general constraints on the equation of state are derived from the general properties expected of dark matter haloes.

\section{The Chaplygin gas and its self-gravitating equilibrium}
\label{sec:equi}

\subsection{Basic properties}
 
The sound speed associated with the equation of state~(\ref{eq:state}) is 
given by
\begin{equation}
c_s^2 = \frac{d p}{d \rho} = A c^2 \alpha \rho^{-\alpha -1}.
\label{eq:sound}
\end{equation} 
As we will see below $A$ takes values that are of the order of the present critical 
density to the power $\alpha +1$. Thus we
can immediately see that the sound speed will tend to zero if $\alpha \rightarrow 0$, for any $\rho$, 
or at high densities (relative to the critical density) if $\alpha$ is large enough. 
This property will be important in what follows as it will determine when pressure 
gradients are important in haloes. 

Using the equation of state, along with those governing 
the evolution of a homogeneous expanding universe, 
one finds that  the density of the homogeneous component in such a universe
evolves with the scale factor as
\begin{equation}
\rho =  \left(A + \frac{B}{a^{3 (1+ \alpha})}\right)^{\frac{1}{1+\alpha}},
\label{eq:chapl}
\end{equation}
where $B$ is another positive constant, and the scale factor is chosen so that 
$a=1$ at the present cosmological epoch.  A Chaplygin gas dominated universe will therefore 
have $\rho \sim 1 /a^3$ for sufficiently small $a$, as in the standard matter dominated case. 
With a present-epoch  energy density  
$\rho (a = 1) =  \rho_c =  \left(A + B\right)^{\frac{1}{1+\alpha}}$,
eventually $\rho \rightarrow A^{\frac{1}{1+\alpha}}$, and the universe is entirely
dark energy dominated.  
It is  thus possible to associate the constant $A$ with the dark energy 
content, and the second term in the brackets of equation~(\ref{eq:chapl}
with a matter contribution. 
This 
fixes the constant $A$ to
\begin{equation}
A = \rho_c^{1+ \alpha} (1 - \Omega_m^{1 + \alpha}),
\label{eq:A}
\end{equation}
where $\Omega_m$ is the present matter contribution to the critical density. 
Alternatively, one may leave $A$ as a free parameter to be determined by best-fitting models 
to observations
(e.g., Wu \& Yu 2007), in which case one obtains 
$ 0.6 \rho_c^{1+ \alpha}  \la  A \la 0.85  \rho_c^{1+ \alpha}$; that is, 
approximately the same values as obtained from equation~(\ref{eq:A}). 
In general,  $A \la \rho_c^{\alpha + 1}$ for a universe that is on the verge 
of being dominated entirely by dark energy. 
In the following we use equation~(\ref{eq:A}) with $\Omega_m =0.3$  
and 
\begin{equation}
\rho_c = \frac{3}{8 \pi G} H_0^2,
\end{equation}
\label{eq:crit}  
to determine $A$. 

Finally, we note that this determination assumed a homogeneous 
Chaplygin gas with no clustering. The effect of clustering on fixing the values 
of the Chaplygin  gas parameters has not been studied in detail. The simple clustering model 
of Avelino et. al. (2014) predicts that  the effective equation of state for the homogeneous 
component of a highly clustered system becomes, as $a \rightarrow 1$, 
$\frac{p_{-}}{\rho_{-}} = -1$ (in units of $c=1$), which implies that $A = \rho_{-}^{\alpha + 1}$. 
At $a=1$, their Fig.~1 suggests that $\rho_{-} \sim 0.75 \rho_c$ if one starts with $90 \%$ 
clustering and with $\alpha =1$. This means that $A \sim 0.6 \rho_c^2$, which is of the same order as 
the estimates above.

\subsection{Hydrostatic equilibrium}

In the standard (CDM) scenario of structure formation, haloes are formed as primordial density 
fluctuations turn around, collapse, and virialise (e.g. Mo, van den Bosch \& White 2010). 
During this process 
the  dynamics  is collisionless and  so  pressure forces  are absent.  Pressure forces are important 
for the dynamics of the gaseous baryonic component, which obeys a conventional equation of state with 
positive pressure  growing with density. This means  that if pressure forces are non-negligible at the initial 
stage of a collapsing gaseous sphere, they can only increase  as this sphere shrinks. 
The situation is  reversed if the sphere is made of unified dark matter, where the magnitude of the pressure decreases with density. In this case, if the pressure forces are negligible at the initial 
stage of the collapse, they will remain so as the system shrinks. 
This property may facilitate the collapse, but  could also render
pressure forces important in the outer regions 
(with small densities) of the final structure that ensues from that collapse.

Before examining the possible effects of pressure gradients on density distributions of outer parts of haloes,  
consider first the hydrostatic equilibrium of a sphere of generalised Chaplygin gas. 
Such an equilibrium can be achieved because, 
although the pressure $p$ is negative, it is more negative for smaller densities; 
it is thus an increasing function  of density, and so its radial derivative is negative when the equilibrium involves a  negative radial density gradient.
Such a pressure gradient can thus act to balance self-gravity.~\footnote{Note that in 
the case of positive pressure decreasing with density, balancing gravity would require a positive density gradient; i.e., 
density increasing with radius, which is not observed in stellar or galactic sysyems.} 
For velocites and densities expected in dark matter haloes, 
the equation for hydrostatic equilibrium reduces to the 
standard Newtonian form, which equates the pressure gradient 
to the gravitational force at radius $r$  (Appendix~\ref{A:norel}):
\begin{equation}
\frac{d p}{d r} = - \rho  \frac{G M}{r^2}.
\label{eq:hydro}
\end{equation} 
where $M = M(< r) = 4 \pi \int_0^r \rho (r) r^2 dr$. Using~(\ref{eq:state}), one can derive
\begin{equation}
\alpha A \frac{d \rho}{dr} = -   \rho^{\alpha +2} \frac{G M}{r^2},
\label{eq:static}
\end{equation} 
which has a power law solution $\rho \sim r^{\frac{-2}{\alpha + 2}}$. For small $\alpha$  
this density profile is similar to the central cusp of an NFW  halo (Navarro, Frenk \& White 1997). And it comes with a rising rotation curve for any positive $\alpha$. 

This is generic; as shown in  Appendix~\ref{A:flatno}, no flat or falling rotation curves 
-- that is, ones associated with density distributions falling as $\rho \sim 1/r^2$ or steeper-- can be 
associated with pressure supported systems with sound speed decreasing with 
density.   Thus, unless  pressure gradients 
are negligible throughout the observable 
regions of galaxies and clusters, or the form of the equation of state  
is modified in the presence of significant density gradients,  
such models are conclusively incompatible with observations. 
Most of the following is concerned with examining the conditions under which 
the first possibility can be realised. 
As pointed out in Appendix~\ref{A:flatno}, the second possibility  does not appear viable.

\section{Transition from CDM-like conditions to hydrostatic pressure support}
\label{sec:trans}

\begin{figure}
\epsfig{file=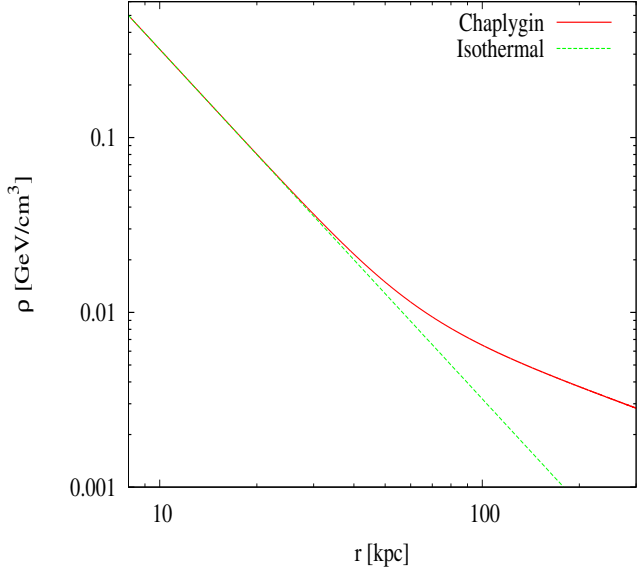,height=8.cm,width=8.9cm,angle=0}
\caption{Outer density profile of a Milky Way like halo
in the presence of Chaplygin gas pressure gradients. The halo properties
are normalised to current estimates of the dark matter density at the 
solar neighborhood. The profile flattens at large radii due to pressure
forces.}
\label{fig:dens}
\end{figure}

\begin{figure}
\epsfig{file=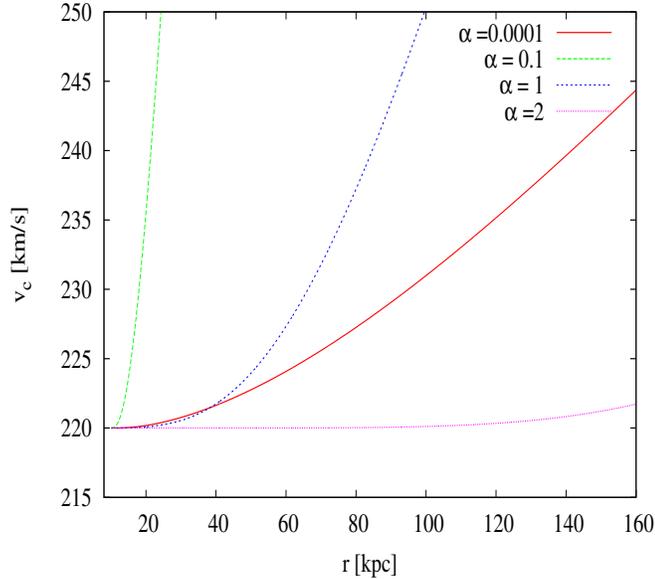,height=8.cm,width=8.9cm,angle=0}
\caption{Outer rotation curves of a Milky Way like galaxy
in the presence of generalised Chaplygin gas pressure, as 
function of the index $\alpha$ of equation~(1). The rotation curves rise
due to the effect of pressure gradients.} 
\label{fig:rot}
\end{figure}

A unified dark  fluid that behaves as pressureless dark matter on galactic scales would be governed
by the Jeans equations.
That is, for  spherical equilibrium structures with isotropic velocity dispersion 
\begin{equation}
\frac{d (\sigma^2 \rho)}{dr} = -  \rho \frac{G M}{r^2}.
\label{eq:Jeans}
\end{equation}
This is the same as equation~(\ref{eq:hydro}), with $\frac{d p}{d r}$
replaced by $\frac{d(\sigma^2 \rho)}{d r}$, where 
where $\sigma = \sigma(r)$ is to be associated with the one dimensional velocity dispersion supporting 
a collisionless equilibrium.  However,  
whereas equation~(\ref{eq:hydro}) has a unique solution, given the equation of state and boundary conditions, equation~(\ref{eq:Jeans}) supports an infinity of density and velocity dispersions pairs that solve it. It is indeed remarkable 
that out of all possible solutions, cosmological halo profiles fall into a highly constrained set, reasonably well 
approximated by the NFW fitting formula. This is not expected to be 
reproduced by the unique solution corresponding to the Chaplygin gas at all radii
(as we saw in the previous section). 

In general,  both the effective pressure, born of the velocity dispersion of 
fluid elements, and the pressure emanating from the equation of state may contribute, so that
\begin{equation}
\frac{d p}{d r}  +   \frac{d(\sigma^2 \rho)}{d r}  = -  \rho \frac{G M}{r^2}.
\label{eq:gen}
\end{equation}
As long as  $\frac{d p}{d r} \ll   \frac{d (\sigma^2 \rho)}{d r}$, this equation reduces to 
the Jeans equation with associated CDM-like behaviour. However, if  
 $\frac{d p}{d r} \ga  \frac{d (\sigma^2 \rho)}{d r}$,
one expects the solution  to 
tend towards that of the hydrostatic equilibrium of the previous section, with its 
rising rotation curve.  
In the regions of flat rotation curve haloes are isothermal, so that 
$\sigma$ is nearly constant. In such regions, the condition for collisionless behaviour becomes 
\begin{equation}
\frac{d p}{d \rho} = c_s^2  \ll \sigma^2;
\label{eq:pcond}
\end{equation}
meaning that, for a given self gravitating mass density distribution,  
the speed of sound of the unified  dark matter fluid  making up the halo has to be much less than the 
equilibrium velocity dispersion that could support the density profile against gravity.
This condition should also hold approximately in the outer parts of NFW haloes where the rotation curves are 
slowly falling with radius.

Flat rotation curves are are associated with density profiles falling as $\sim 1/^2$, and in a collsionless systems 
these are associated with constant velocity dispersion $\sigma$. In a system made of unified dark matter, 
additional  pressure forces may contribute, corrupting the $\sim 1/r^2$ 'isothermal' profile; since, as we saw 
in the presbious section, when this pressure dominates, the density profiles fall less steeply with radius. 
To check if flat rotation curves remain  uncorrupted  over observable 
radii, one  can solve equation~(\ref{eq:gen}) with constant $\sigma$ and look  for deviations from an isothermal 
profile, born of  non-negligible pressure gradient~\footnote{Little is known of the dynamics of the transition 
from collisionless to hydrodynamic behaviour in a Chaplygin gas, it seems likely however that the velocity 
dispersion would decrease considerably as pressure forces become significant (as in standard hydrodynamics transition). 
Hence the assumption of constant $\sigma$ may actually 
{\em underestimate} the density (and overestimate the radius) 
at which pressure gradients should have significant effect on the rotation curves.}. 
{If  $\sigma$ is constant and $p$ is solely a function of density, one can differentiate~(\ref{eq:gen}) with 
respect to $r$, while making use of the relation $dM/dr = 4 \pi r^2  \rho$, to obtain
\begin{equation}
\frac{d^2 \rho}{dr^2}  -   \frac{X}{Y} \left(\frac{d \rho}{dr}\right)^2 + \frac{2}{r} \frac{d \rho}{dr} 
= - \frac{4}{Y}  \pi G \rho^2. 
\label{eq:iso}
\end{equation}
Here
%
$X = \frac{d^2 p}{d \rho^2} - \frac{\sigma^2}{\rho^2} - \frac{1}{\rho} \frac{d p}{d \rho}$
%
and
$Y=  \frac{d p}{d \rho} + \sigma^2$.
For a generalised Chaplygin gas  $X = \frac{\alpha (\alpha + 2)}{\rho^{\alpha+2}} A c^2 + \frac{\sigma^2}{\rho}$ and 
$Y= \alpha A c^2 \rho^{-\alpha + 1 } + \sigma^2$.  

This procedure is similar to that used in obtaining the Lane-Embden equation for stellar polytropes, though
we skip here the step of transforming to dimnensionless variables here, keeping the physical quantities which, for illustration, will be directly compared with the specific case of a  Milky Way like galaxy counterparts.}
Equation~(\ref{eq:iso}) was thus integrated while fixing 
the  parameters to values relevant to a Milky Way like galaxy. 
In the absence of pressure forces, the halo is assumed to be a singular isothermal sphere, with 
$\rho (r) = \frac{\sigma^2}{2 \pi G r^2}$, and $\sigma$ chosen so that
the circular rotation speed ($v_c^2 = 2 \sigma^2$) is 220~km/s.
The corresponding density  at solar radius
 is consistent with the  local dark matter density inferred from observations and  expected from 
standard CDM halo profiles  (e.g., Salucci et. al 2010; Bovy \& Tremaine 2012);
if slightly on the higher end, as the asymptotic rotation speed is assumed here to arise 
entirely from the gravity of the dark matter component (figures~16 and~17 of Xu et. al. 2008, for example, 
suggest that a baryon contribution to the rotation speed of $\sim 15-20  \%$ at 60~kpc, which translates in a factor of about 1.4 
in density).   At $r \sim 10$~kpc, the pressure 
forces are negligible, so that one can use the density for the isothermal sphere  
 (and the associated radial derivative)  to define the  initial value problem and 
integrate   equation~(\ref{eq:iso}) outwards.
Note that since equation~(\ref{eq:iso})  
is local, the fact that a singular isothermal sphere may not correctly reproduce the 
dark matter mass distribution at very small or large radii is not important, as long we are interested 
in a region where a collisionless dark matter density should fall off as $\sim 1/r^2$.

The results are shown in Fig.~\ref{fig:dens} for the Chaplygin gas ($\alpha = 1$). 
Condition~(\ref{eq:pcond})  starts to be violated at 
$r \sim 50$~kpc; accordingly, the density profile  starts deviating from isothermal as expected.  
The corresponding rotation curves are shown in Fig.~\ref{fig:rot} for several values 
of $\alpha$. As can be seen, unless $\alpha$ is very small 
(associated with a $\Lambda$CDM cosmology) or greater than one (and thus can correspond 
to superluminal sound speed), the rotation curve sharply deviates from flatness, rising steeply
at radii significantly smaller than the expected Milky Way halo virial radius ($\sim 200$~kpc).
This  is a consequence of the form of the sound speed as defined in equation~(\ref{eq:sound}); the multiplicative 
factor $\alpha$ ensures a small sound speed is  relatively small at all densities for small $\alpha$,
while the sound speed decreases steeply with density for large enough $\alpha$.  

For the Milky way there is no evidence that the rotation curve takes to rising at such radii. Indeed, if 
anything,  all the evidence suggests the 
contrary (Xue et. al. 2008; Bhattacharjee, Chaudhury \& Kundu 2014). 
There are also many galaxies where direct measurements of the rotation curve 
exclude the possibility, while still exhibiting $c_s^2/\sigma^2 >1$.  
Some examples are shown on Table~1. (As in the case of the Milky Way above,  
the rotation curves are assumed to be entirely dominated by dark matter at the largest radii probed.
We thus use $\sigma^2 = \frac{1}{2} v_c^2$, where $v_c$ is the observed circular velocity at the maximum
radius probed.  The sound speed is calculated from equation~\ref{eq:sound}, using the expected dark fluid 
density of the isothermal sphere associated with $\sigma$.
Including a baryonic contribution would decrease the needed dark sector density, decreasing  $\sigma$ and further increasing  $c_s$).

\begin{table}
\begin{tabular}{ |p{2cm}||p{1cm}|p{1cm}|p{1cm}||p{1cm}|  }
\hline
Object &  $R_{\rm max}$ & $\sigma$  &  $c_s$ & $c_s^2/\sigma^2$ \\
\hline
NGC 3741  & 7  &   32    &  51  &   2.5  \\

Malin 1 &  88  &  151    &  352  &  5.4  \\

U1230   & 35  &   72    &   244 &  11.5   \\
 
U5005  &   28  &  70    & 166 & 5.6 \\

U5750  & 22 &  56  & 161  &  8.3\\
 
NGC5055 & 48 &  120    & 168 & 2 \\

UGC 11852 &  90  &  117    &  626 & 28.6    \\ 

N3198 &  44  &  106  &  181 & 2.9    \\

D154  & 10    & 29     & 125 & 18.6 \\

 \hline
\end{tabular}
\caption{Examples of the velocity dispersions $\sigma$, required to support an isothermal 
halo  profile with flat rotation curve, exceeded by the Cahplygin gas ($\alpha =1$) sound speed $c_s$ 
at maximum radius, $R_{\rm max}$, probed observationally. 
Data for NGC 3741 is taken from Gentile et. al. (2006); Malin 1, from 
Lelli, Fraternali \&  Sancisi (2010);  U1230 and U5005  from de Blok \& McGaugh (1997); 
U5750 from  McGaugh, Rubin \& de Blok (2001);
NGC5055, from de Blok et. al. (2008); UGC 11852, from Noordermeer et al.(2007);
N3198 and D154, from Bottema \& Pestana (2015)}
\end{table}

\section{Clusters of Chaplygin gas?}
\label{sec:clus}

The radially extended hot gas in clusters of galaxies allows for the  probing of the gravitational potential up to large radii. 
When the density profile is decomposed,  the dark halo density contribution can be reasonably 
fit using the standard NFW profile. 
Given this,  the pressure forces associated with Chaplygin gas halo need to be 
negligible, contributing insignificantly to the support against self gravity out to the maximum radii probed
(so that the configuration behaves as a collisionless equilibrium that can be fit by NFW). 
This section is concerned with checking whether this is true for a sample of galaxy clusters. 
This will be done by using NFW fits inferred from observations, then calculating the  pressure force associated with the  
NFW density at the maximal observed radius, then comparing it with the force of self gravity of the inferred halo at that radius. 
{Note that the NFW is employed here as a useful parametrisation (as it has been extensively used 
in fitting cluster profiles); the crucial point being that it fits steeply decreasing outer density
profiles of observed clusters, while the profiles associated with a unified  dark  fluid with significant pressure gradient 
do not (as they fall off too slowly with radius as seen in the previous sections). Any profile that fits the 
observed mass distribution  equally well would give similar constraints, within the errors 
associated with the particular empirical fit,  as those deduced below.}

The NFW profile is given by
\begin{equation}
\rho_{\rm NFW} = \frac{\rho_0}{\frac{r}{r_s} (1 + \frac{r}{r_s})^2},
\label{eq:NFW}
\end{equation} 
where $\rho_0$  is a characteristic density and  $r_s$  the radial scale length. 
For a halo concentration parameter $c_{\rm vir}=r_{\rm vir}/r_s$
\begin{equation}
\rho_0 = \frac{M_{\rm vir}}{4 \pi r_s^3 F (c_{\rm vir})}, 
\label{eq:vir}
\end{equation}
where  the function in the denominator is given by
\begin{equation}
F (x) = \ln (1+ x) - \frac{x}{1 + x}.
\end{equation}
Defining 
$x = r/r_s$, the pressure gradient of a generalised Chaplygin gas  with this density distribution is given by  
\begin{equation}
F_P = \frac{d p}{d r} = - A c^2 \frac{\alpha}{r_s} \rho_0^{-\alpha} (1 + 3 x) x^{\alpha -1 } (1+x)^{2 \alpha -1};
\end{equation}
while the self gravity term for such a  configuration takes the form 
\begin{equation}
 F_G = - \rho \frac{G M}{r^2} =  - 4 \pi G \rho_0^2  r_s  \frac{F(x)}{x^3 (1+x)^2}.
\end{equation}  
\begin{figure}
\epsfig{file=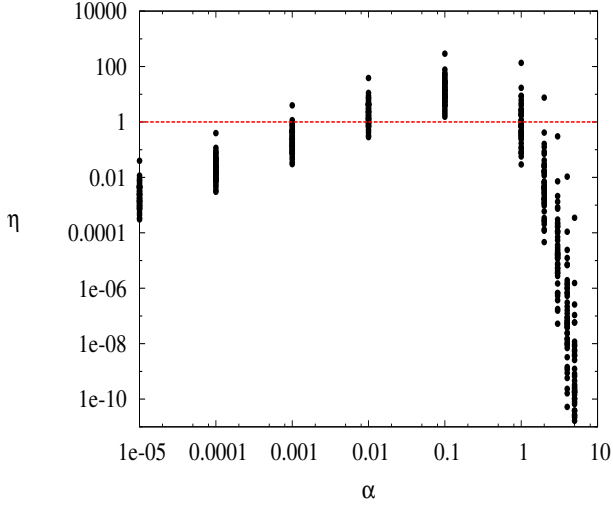,height=7.cm,width=8.5cm,angle=0}
\caption{Ratio of the expected generalised 
chaplygin gas pressure forces to gravitational forces  (equation~\ref{eq:rat})
for NFW density fits recovered by Ettorri et. al (2010) for 
sample of galaxy clusters. The ratios are plotted at the maximum radii probed
and for various values of the index $\alpha$ of equation~1.}
\label{fig:eta}
\end{figure}

\begin{figure}
\epsfig{file=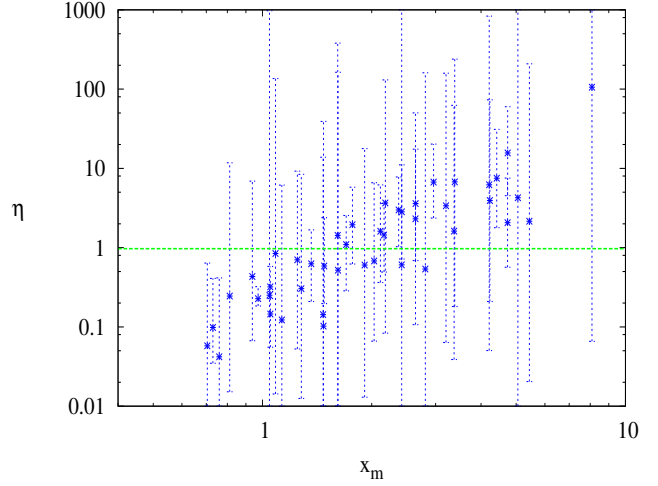,height=7.cm,width=8.8cm,angle=0}
\epsfig{file=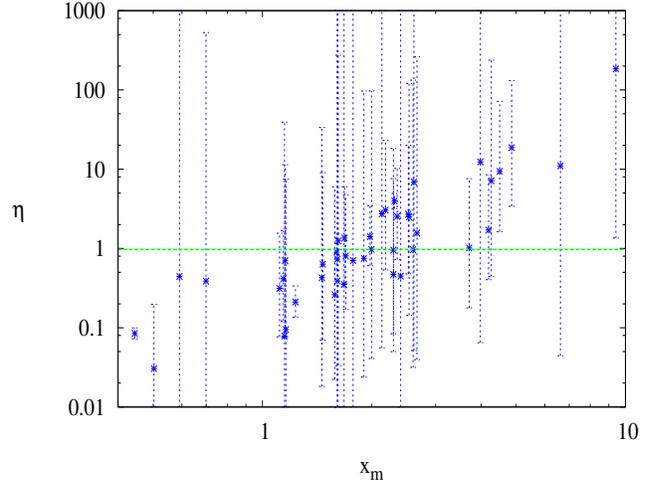,height=7.cm,width=8.8cm,angle=0}
\caption{Ratio of the expected  
chaplygin gas ($\alpha =1$)pressure forces to gravitational forces  (equation~\ref{eq:rat})
for NFW density fits recovered by Ettorri et. al (2010), as function of the 
maximum radii probed normalised to the scale radius $r_s$. The error bars 
represent the maximum possible errors resulting from uncertainities in both the measured
values of both $r_s$ and concentration parameter $c_{\rm vir}$.}
\label{fig:xm}
\end{figure}

\begin{figure}
\epsfig{file=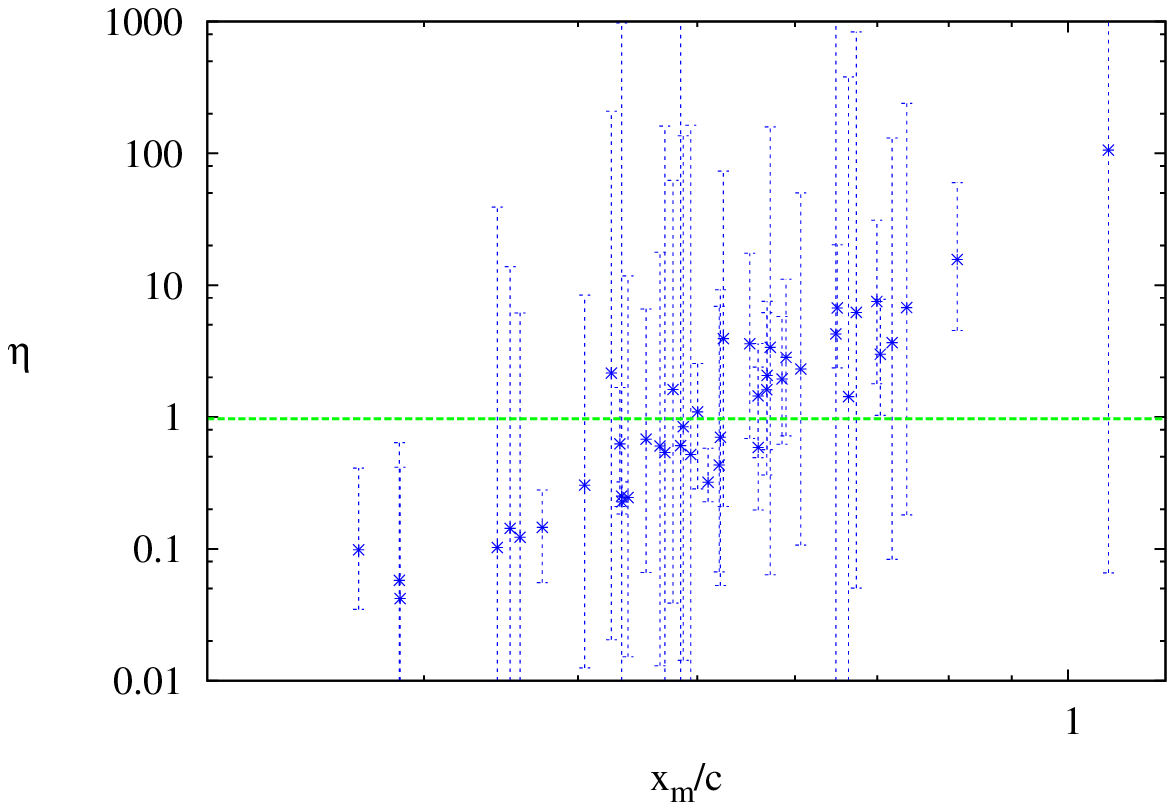,height=7.cm,width=8.8cm,angle=0}
\epsfig{file=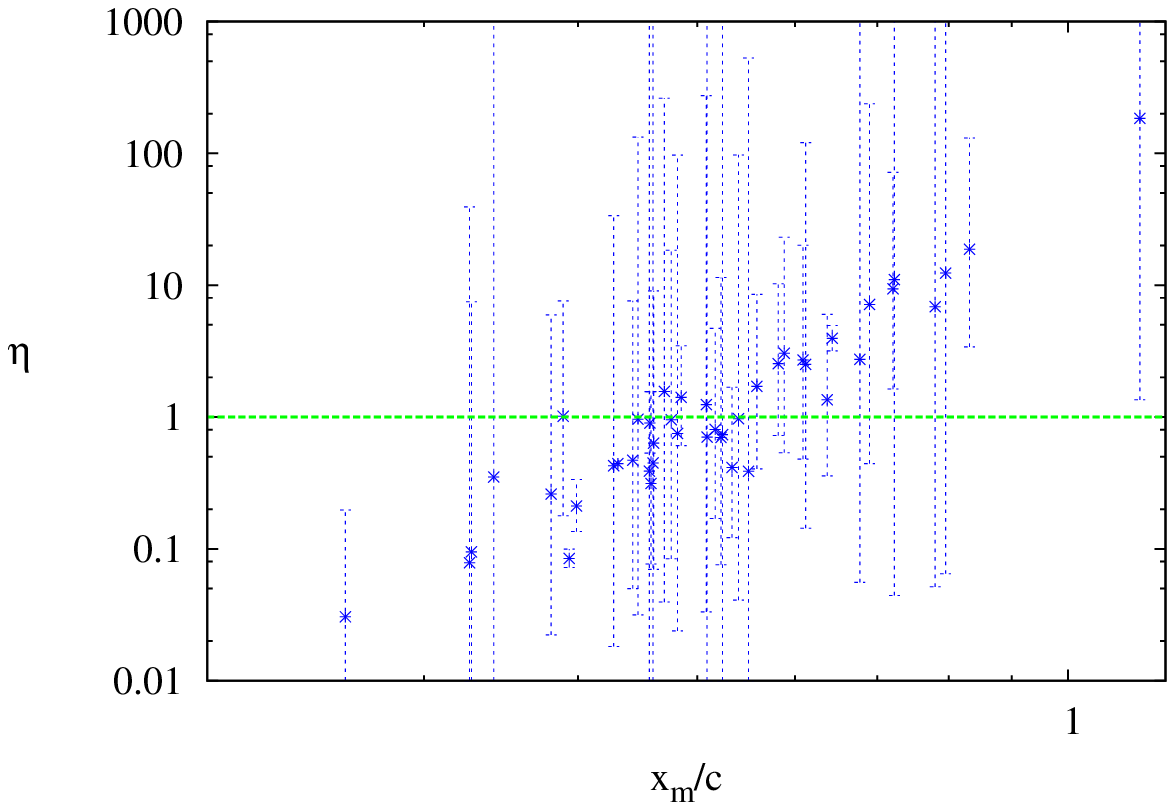,height=7.cm,width=8.8cm,angle=0}
\caption{Same as in Fig.~\ref{fig:xm} but with maximum radii normalised
to the virial radius $r_{\rm vir} = c r_s$.}
\label{fig:xmc}
\end{figure}

In this context, we define the ratio 
$\eta = \frac{F_P}{F_G}$
At large radiii,  $\rho_{\rm NFW} \sim 1/r^2$ -- $1/r^3$, which cannot be maintained 
if the pressure gradient is non-negligible, as we saw in the previous sections.  Therefore, it is necessary that,
at the maximum radius probed by the observations,
\begin{equation}
\eta = \frac{\alpha A c^2} {4 \pi G \rho_0^{2 + \alpha}} \frac{x_m^{2 + \alpha} (1 + x_m)^{1 + 2 \alpha} (1 + 3 x_m)}
{r_s^2 F (x_m)} \ll 1,
\label{eq:rat}
\end{equation}
where $x_m = r_{\rm max}/r_s$,  $r_{\rm max}$ being the maximal radius probed by observations.
We now wish to test where this is the case for actual clusters of galaxies. For that purpose we use the results 
of Ettori et. al. (2010),
averaging over the two methods they employed for deducing the NFW parameter and setting 
$x_m = R_{\rm sp}/r_s$  and $x_m = R_{\rm xsp}/r_s$, as given in their Table~2.

The results are shown in Fig.~\ref{fig:eta}.  As can be seen, unless the index $\alpha$ is very small  ($ \la 0.0001$)
or large enough ($\alpha \ga 2$), the ratio $\eta$ 
can be of order one or larger. This is agreement with the results of the previous section, and also of linear 
analysis. As opposed to the linear case, however, the degree of clustering of the large scale Chaplygin gas 
does not appear to be important, since the inferred value of $A$ does not seem to vary much with clustering
(c.f. Section~2). 

As will be seen in Fig.~\ref{fig:eta}, for each value of $\alpha$ there is significant scatter in $\eta$. This includes intrinsic scatter 
in the values of the inferred halo parameters; significant errors entering in the estimation of those parameter; 
as well as systematics due to the variation of the maximum radius probed by the observations. 
We examine in more detail the important case of $\alpha = 1$
in Fig.~\ref{fig:xm}~and~\ref{fig:xmc}.  From those figures one infers that  there is systematic increase in $\eta$ for clusters that are observed  out to larger radii. It is therefore safe to assume that-- 
despite the large errors in $\eta$, stemming from the collusion of errors in estimates of NFW halo
scale lengths and concentration parameters-- 
the constraints on $\alpha$ are in fact  rendered conservative by the limitations 
of maximum radii probed.  Probing to larger radii is likely to systematically increase the deduced values of $\eta$, further
constraining the model. In the next section we show that this is the case, as  tighter constraints are inferred from 
NFW halo properties at the virial radius.

\section{Unified dark matter and halo parameters}
\label{sec:gen}

{Direct observations of the internal dynamics of galaxies and clusters usually fall well short of 
the expected virial radii of the haloes believed to envelop these systems. Nevertheless, weak lensing
fields are also well fit by assuming that collapsed structures follow NFW-like density profiles
 (e.g., Mandelbaum 2006; Oguri \& Hamana 2011); if anything the fits are in fact better when these are truncated beyond the virial radius, so as  
to fall faster than the $1/r^3$ behaviour associated with an outer NFW profile. This is again 
incompatible with the slowly falling density profiles of pressure supported unified dark matter 
fluids.  In this section we obtain  further constraints, based on properties of NFW haloes 
derived from simulations, by requiring that the unified dark fluid pressure gradient be small up
to the virial radius.}

In the most general terms, in order for a unified dark matter fluid
to behave as CDM, one requires that $\frac{d p}{d r} \ll  \frac{d (\rho \sigma^2)}{dr}$.
For an NFW halo and from $r \ga   r_s$ out to the virial radius, $\sigma^2$ is a  slowly varying function of $r$ 
and 
so the above condition reduces to that of equation~(\ref{eq:pcond}),
meaning that the speed of sound must be much less that the 
isotropic equilibrium velocity dispersion that  could support the density profile against gravity. 
At large radii, the velocity dispersion of an NFW halo tends to an
asymptotic value such that  $\sigma^2 = \frac{1}{4} v_c^2$ 
(e.g.,  El-Zant 2006, Appendix B), so that 
\begin{equation}
\sigma^2 =  \pi  \rho_0 r_s^2  G  \frac{F(x)}{x}.
\end{equation}
We can eliminate $r_s$ by using~(\ref{eq:vir}). In addition $c_{\rm vir}$ and $M_{\rm vir}$ are 
found to correlate in numerical simulations, namely through power law relations of the form
\begin{equation}
c_{\rm vir} = A_h \left(\frac{M_{\rm vir}}{B_h M_{\odot}}\right)^{-b}.
\end{equation}
The logarithmic slope $b$ is found to be of the order of $0.1$, and the parameters 
$A_h$ and $B_h$ take values such that $c \sim 10$ for 
a halo of $10^{12}$ solar masses. 
For example Klypin, Trujillo-Gomez, \& Primack  (2011)
obtain  $A_h = 9.6$ and $b =0.075$ for $B_h =10^{12} h^{-1}$. 
(There is also an apparent 
upturn in the mass concentration relation at high masses, though it has been argued 
that this is due to the inclusion of non-relaxed systems; 
Ludlow et. al. 2012). 

Using these relations one finds that
\begin{equation}
\sigma^2 = C^2 
\left(\frac{\rho_c}{\rho_0}\right)^{\epsilon -1} \frac{F(x)}{x  F (c_{\rm vir})^\epsilon}
\end{equation} 
where 
\begin{equation}
\epsilon = \frac{2 (1+ 3 b)}{9 b},
\end{equation}
\begin{equation}
C^2 =  G (\pi \rho_c)^{1/3} 
\left(\frac{A_h^{1/b} B M_\odot}{4}\right)^{2/3} 
\left(\frac{\Delta}{3}\right)^{\epsilon -2/3},
\end{equation}
and $\Delta$ is the halo overdensity within the 
virial radius, such that (at $z=0$)
$M_{\rm vir} = 4 \pi \rho_c \frac{\Delta}{3} c_{\rm vir}^3 r_s^3$ and 
$c_{\rm vir}^3 = 3 \rho_0 \frac{F(c_{\rm vir})}{\rho_c \Delta}$.  
For $\Delta= 200$, and values of $A_h$, $B_h$ and $b$ mentioned above,
$C^2$ is of order $10^{14} ({\rm km/s})^2$.

At the virial radius
\begin{equation}
\sigma^2 = \frac{C^2}{c_{\rm vir} F (c_{\rm vir})^{\epsilon-1}}.
\left(\frac{\rho_c}{\rho_0}\right)^{\epsilon -1} 
\end{equation} 
This is a general constraint that the equation of state of any unified dark matter 
fluid should obey. That is, the sound speed should either be small everywhere
or decrease sufficiently with density so that the above relation is satisfied at the virial 
radius. 
For example,for the  generalised Chaplygin gas with an NFW profile 
the sound speed is 
\begin{equation}
c_s^2 \approx \alpha c^2 \left(\frac{\rho_c}{\rho_0}\right)^{\alpha + 1} 
c_{\rm vir}^{3 (\alpha +1)},
\end{equation}   
(where we have assumed large enough $c_{\rm vir}$ so that $c_{\rm vir} +1 \rightarrow c_{\rm vir}$
and $A \approx \rho_c^2$). 
For a generalised Chaplygin gas to behave as pressureless dark matter 
with an NFW profile up to the virial radius thus requires 
\begin{equation} 
\frac{c_s^2}{\sigma^2} \approx \frac{c^2}{C^2} \alpha \left(\frac{3 F(c_{\rm vir})}{\Delta}\right)^{\alpha + 2 - \epsilon} 
c_{\rm vir}^{3 \epsilon -2} \ll 1, 
\end{equation}
where we have used $\rho_0/\rho_c = \frac{\Delta}{3} \frac{c_{\rm vir}^3}{F(c_{\rm vir})}$.  
For $b =0.1$, $\epsilon = 2.9$, and for a Milky Way like galaxy 
$c_{\rm vir} \approx 10$, so that $\alpha$ needs to be either very small or 
significantly larger than $- (2 - \epsilon)$ for the above inequality to be satisfied. For example, 
for $c_{\rm vir}$ one needs either $\alpha < 2  \times 10^{-6}$ or $\alpha > 3.8$ for the ratio to be less than one;, 
for $c_{\rm vir} =30$
the index is further constrained to $\alpha <  10^{-9}$ or $\alpha > 6.7$.

\section{Conclusion}

In unified dark matter models both dark energy and dark matter are 
reproduced by one component, which can behave as pressureless 
dark matter at high density and as negative pressure dark energy at low densities. 
In a cosmological context, past the radiation era, 
this implies a phase of matter domination followed by dark energy driven 
expansion.

 In terms of halo structure, the unified dark matter scenario also has its appeal, in principle: 
bound object, if they form, should have density high enough so that the dark  
fluid they are made of behaves as pressureless dark matter, of the sort that 
the haloes harbouring  galaxies are believed to be made of. 
Nevertheless, the densities of dark matter haloes is not exceedingly large; the average mass within
the virial radius is by definition only $\sim 200$ times larger than the ambient density of the universe, 
it is smaller still in the outer regions. One may therefore ask if the pressure forces associated with the 
unified dark matter can have significant  effect on the structure of such haloes  at observable radii, 
and what that effect, if present, may be. 

In the prototypical case of the generalised Chaplygin gas it was shown 
that -- in order to reproduce the observed outer rotation curves in galaxies and inferred outer 
density profiles in clusters --
the index $\alpha$ of equation~(\ref{eq:state}) should be  constrained so that either
$\alpha \ga 2$ or  $\alpha \la 0.0001$. {Cosmological models with small $\alpha$ 
 are quite close to $\Lambda$CDM. The sound speed in large $\alpha$ models can be superluminal, though it has been argued that this does not necessarily violate causality (Gorini et. al. 2008a).    
Large $\alpha$ models are also in principle observationally distinguishable from  $\Lambda$CDM, as they 
have the interesting imprint of describing a background evolution that transitions rather abruptly 
from a matter dominated to a de Stter phase (Piatella 2010).}  

The bounds derived here  are similar to those obtained from studying the linear clustering regime (e.g., Gorini et. al. 2008), though in that  latter case it appears that the constraints may be circumvented by invoking 
the effect of nonlinear clustering on the effective 
equation of state (Avelino et. al. 2014).   This effect has not been studied in detail,
though the simplest model (described in the aforementioned reference) 
does not imply modifications that would affect the results presented here significantly
(cf. Section~2.1). 

{The discrepancies, leading to the 
aforementioned constraints,  are  generic; they stem from the observational requirement that 
the outer density profile of galaxies and clusters fall off at least
as $\sim 1/r^2$, while those unified dark matter systems have significantly shallower profiles
when the pressure dominates.    Indeed, for  a generalised Chaplygin gas configuration with significant pressure gradient, the density falls as
$1/r$ or slower
(Section~2.2).}  In general, it can be shown that, 
for any dark fluid with barotropic equation of state and sound speed decreasing with density, 
profiles  associated with flat 
or falling rotation curves -- and therefore also a density distribution characteristic 
of the outer parts of clusters -- cannot be supported when the unified dark fluid 
pressure forces are dominant (Appendix~B). That the sound  speed decreases with increasing 
density is a necessary condition if the fluid is to behave as collsionless dark matter at high density. 

Requiring that unified dark matter haloes conform to NFW profiles up to the virial radius 
leads to further, tighter, constraints on the sound speed and associated equation of state.
{Although the range of empirical dynamical evidence prsented here mostly falls well short of the expected 
virial radii of the haloes of the systems examined, weak lensing signals, which probe the density 
distribution to mauch larger radii,  are well modelled by invoking NFW-like 
halo density distributions  (e.g., Mandelbaum 2006; Oguri \& Hamana 2011) . The NFW profile serves as a useful paramtrisation of the  density profile of 
the gravitationally bound component of the mass distribution when fitting the lensing data; it
predicts an outer density profile falling as $\sim 1/r^3$. if anything, better fits to the lensing data are achieved by truncating the halo profiles used  so that the 
density falls off far faster than NFW 
beyond the virial radius. This is in stark contradiction with the slowly falling density profiles 
we find  when a pressure contribution associated with a unified dark fluid 
is significant in the outer regions of self
gravitating configurations. These shallow profiles should persist well beyond the virial radius typical of a CDM I
halo, in fact they should continue until the unified dark matter fluid composing the 'halo' reaches 
pressure equilibrium with the ambient homogeneous 
background representing the dark energy. 
Such a situation would  appear to be in clear contradiction with the lensing data. 
To fit that data one needs 
a rapidly falling desnsity distribution around in the outer parts of self gravitating 
configurations.}

\section*{Acknowledgments}
I would like to thank Gary Tupper for useful discussions and 
Adel Awad form commenting on the manuscript.

\appendix

\section{The Newtonian limit in a generalised Chaplygin gas halo}
\label{A:norel}
The equation for hydrostatic equilibrium of a relativistic spherical system composed of a perfect fluid
with density $\rho$, pressure $p$ and mass enclosed within radius $M = M (< r)$ enclosed within radius $r$ is   
(e.g. Weinberg 1972)
\begin{equation}
\frac{dp}{dr} = - \frac{G}{r^2} (\rho + p/c^2) \left(M  +  \frac{4 \pi r^3 p}{c^2}\right)
\left(1 -  \frac{2 G M}{r c^2}\right)^{-1}.
\end{equation}
In the case of generalised  Chaplygin gas with equation of state of the form~(\ref{eq:state}), 
the rich variety of behaviour in the relativistic self gravitating regime has been described for example by 
Neven, Tupper \& Viollier (2006) and by Gorini et. al. (2008b; 2009). In this paper we consider 
the Newtonian limit; which, as we will see, should apply  to  configurations with properties akin to those of 
dark matter haloes.  

In a dark halo  the circular velocity $v_c$ is much smaller than the speed of light $c$, so that 
\begin{equation}
\frac{2 G M}{r c^2} = 2 \frac{v_c^2}{c^2} \ll 1.
\end{equation}
Also, using equation~(\ref{eq:state}) and~(\ref{eq:A}), one finds that 
\begin{equation}
\frac{-p}{\rho c^2} \approx  \left(\frac{\rho_c}{\rho}\right)^{1 + \alpha} \ll 1.,
\end{equation}
since the density within a halo is always much larger than the critical 
density of the universe. 
Finally,
\begin{equation}
\frac{4 \pi r^3 p}{M c^2} \approx \frac{4 \pi r^3 \rho_c}{M}  \left(\frac{\rho_c}{\rho}\right)^{\alpha}.
\end{equation}
Since $4 \pi r^3 \rho_c$ is of the same order of the mass of homogeneous 
distribution with density $\rho_c$ enclosed inside a sphere of radius $r$, this is 
much smaller than the actual mass $M(r)$ inside such a sphere. The term in brackets
is also small (as above), so that 
hand side is also much smaller than one.

Under these conditions the condition for hydrostatic equilibrium reduces to the usual
Newtonian equation
\begin{equation}
\frac{d p}{d r} = - \rho \frac{G M (<r)}{r^2}
\end{equation}

\section{The absence of flat or falling rotation curves}
\label{A:flatno}

It is possible to obtain flat rotation curves  by deriving the equations of motion of the Chaplygin gas  
from a scalar field  while  assuming that the latter possesses  space-like gradients in the collapsed (self gravitating) 
phase, which is thus characterised by an accordingly modified equation of state  (e.g., Armendariz-Picon \& Lim 2005). 
However this comes at a hefty price. The gas is no longer an isotropic perfect
fluid, the corresponding density distribution has poles and is undefined in  the central regions
of the self gravitating system, and it general no longer satisfies the positive definite energy 
condition (Diez-Tejedor  \& Gonzalez-Morales 2013).   In this paper, as in the studies cited  in the previous appendix, it is assumed
that the equation of state remains unaltered in the collapsed phase. It will now 
be shown that, in this case, no flat or falling rotation curves can be obtained for a self
gravitating system with non-negligible pressure gradient, if the sound speed is a decreasing 
function of density. 
 
Cosmology requires that the unified dark fluid behaves 
as a pressureless matter component prior to dark energy 
domination. If the equation of state remains unmodified this  
implies that, at high enough density, that the unified dark matter fluid 
composing the halo behaves as pressureless dark matter 
at small radii. Nevertheless, pressure forces can become 
dominant in the outer, low density, regions
of the halo. As the rest of this paper shows, this will be the case, 
for example, for a generalised
Chaplygin gas, unless its index $\alpha$ is very small or significantly larger than 
one. We will now show that in such cases it is not possible to sustain 
flat or falling outer rotation curves. 

When the pressure dominates the support against gravity,
the circular speed ($v_c^2 = G M/r$) in a halo 
with pressure obeying the equation an
equation of  state $p = p(\rho)$ and associated 
sound speed $c_s^2 = \frac{d p}{df \rho}$,  is 
(using equation~\ref{eq:hydro})
\begin{equation}
 v_c^2 =  - c_s^2 \frac{r}{\rho} \frac{d \rho} {d  r} 
= - c_s^2 \frac{d \ln \rho}{d \ln r}.
\end{equation}  
Assuming a real sound speed ($c_s^2 \ge 0$) therefore implies
that the density cannot increase with radius. 
From the above we also have 
\begin{equation}
\frac{d}{d r} \left(\frac{v_c^2}{c_s^2}\right) = - \frac{d}{d r} \left(\frac{d \ln \rho} {d \ln r} \right).
\label{eq:logs}
\end{equation}
Now, assume we have a flat rotation curve up to some radius $r_0$ (so as to fit observations), 
the associated logarithmic slope around $r_0$  
should then be $-2$ (so that $\rho \sim 1/r^2$, $M (< r) \sim r$ and $v_c^2$ is constant).
If the rotation curve is to stay flat, or fall,  the logarithmic slope should be 
$\le -2$ for $r > r_0$. In this case,  the right hand side of equation~(\ref{eq:logs})
is non-negative.  On the other hand, for a flat or falling rotation curve the left hand side 
can only be positive if $c_s^2$ decreases with radius.
But if the unified dark matter fluid is to behave as CDM for high density 
$c_s^2$ must decrease with density, 
and therefore {\em increase} with radius. Thus it is impossible to have 
flat or falling rotation curves in this case.

\end{document}